\documentclass[oldversion]{aa}
\usepackage{times,graphicx,amssymb,lscape}
\usepackage{rotating}
\usepackage{natbib}
\bibpunct{(}{)}{;}{a}{}{,}  

\begin{document}

\title{High-Frequency Oscillations in a Solar Active Region observed with the {\sc{rapid dual imager}}}

\author{D.B. Jess\inst{1,2}
        \and
        A. Andi\'{c}\inst{1}
	\and
        M. Mathioudakis\inst{1} 
        \and
        D.S. Bloomfield\inst{3}
        \and
        F.P. Keenan\inst{1}
        }

\institute{ Astrophysics Research Centre, School of Mathematics and Physics, Queen's University, Belfast, BT7~1NN, 
Northern Ireland, U.K.
\and
NASA Goddard Space Flight Center, Solar Physics Laboratory, Code 612.1, Greenbelt, MD 20771, USA
\and
Max-Planck-Institut f\"{u}r Sonnensystemforschung, Max-Planck-Str. 2, 37191 Katlenburg-Lindau, Germany
}

\offprints{D.B. Jess, \email{djess01@qub.ac.uk}}

\date{Received 22 January 2007 / Accepted 16 July 2007}

\abstract{
High-cadence, synchronized, multiwavelength optical observations of a solar active region (NOAA 10794) are presented. The data were 
obtained with the Dunn  Solar Telescope at the National 
Solar Observatory/Sacramento Peak using a newly developed camera system : the {\sc{rapid dual imager}}. Wavelet analysis is undertaken 
to search for intensity related oscillatory signatures, and 
periodicities ranging from 20 to 370~s are found with significance levels exceeding 95\%. Observations in the H-$\alpha$ blue 
wing show more penumbral oscillatory phenomena when compared to simultaneous G-band observations. The H-$\alpha$ oscillations are 
interpreted as the signatures of plasma motions with a mean velocity of 20~km/s. The strong oscillatory power over 
H-$\alpha$ blue-wing and G-band penumbral bright grains is an indication 
of the Evershed flow with frequencies higher than previously reported.

}

\authorrunning{D.B. Jess et~al.}
\titlerunning{High-Frequency Solar Oscillations}

\maketitle 
 
\section{Introduction}
\label{intro}

Since the discovery of solar oscillations in the 1960s (Leighton 1960), and their subsequent confirmation by 
Deubner \cite{Deu75}, there has been a multitude of observational evidence presented verifying the existence of oscillations 
in the solar atmosphere (Stein \& Leibacher 1974). Oscillations have been suggested as candidates to explain one of 
the main unanswered questions in solar physics --  {\it Why is the outer solar atmosphere hotter than its surface?} Acoustic 
oscillations as a mechanism to support atmospheric heating in the form of wave dissipation was postulated by Schwarzschild 
\cite{Sch48} and Biermann \cite{Bie48}. Theory suggests that this form of magnetohydrodynamic (MHD) wave can either propagate 
upwards from the lower solar atmosphere or be induced in active regions by reconnection events (Hollweg 1981). 
Early work concerned with oscillatory phenomena in highly magnetic structures ({\it{e.g.}} sunspots), have validated the detection of 
low-frequency oscillations which are a response of the umbral photosphere to the 
5-minute p-mode global oscillations (see review by Lites~1992). Furthermore, Bloomfield~et~al.~\cite{Blo07} have examined wave modes 
within sunspot penumbra and have determined that modified p-mode waves exhibit the best agreement with current observations. Similarly, 
Marcu~\&~Ballai~\cite{Mar05} have studied oscillations with reference to the propagation of compressional MHD waves in penumbral 
filamentary structures in the photosphere.

Extensive analyses of low-frequency ($\le~3.6$~mHz) oscillations has shown evidence for mode coupling in the lower solar atmosphere 
(McAteer~et~al.~2003, Bloomfield~et~al.~2004). Higher-frequency acoustic oscillations ($3.3$~-~$33$~mHz) have been detected in the 
chromosphere (White \& Athay 1979a, 1979b and Athay \& White 1978, 1979a, 1979b), but it was found that the total energy flux 
of the oscillations was two orders of magnitude lower than that required to balance the radiative losses from the chromosphere. 
Further work on high-frequency oscillations has been performed by Williams~et~al.~(2001, 2002), who interpret 6~s intensity 
oscillations as longitudinal magneto-acoustic waves in an active region coronal loop, and by Andi\'{c}~\cite{And07} who 
investigated the radial component of chromospheric oscillations with periodicities as short as 45~s.

Following analysis of {\sc{trace}} data, Fossum \& Carlsson \cite{Fos05a} were unable to detect sufficient power in high-frequency waves 
and concluded  that these waves cannot constitute the dominant heating mechanism of the chromosphere. However, this study is limited by 
the cadence {\sc{trace}} can achieve and the onboard filter transmissions (Fossum \& Carlsson 2005b).  
This means that physically small oscillation sites with short coherence lengths may be smeared out by the coarse sampling. 
In addition this method may overlook dynamic patterns created on spatial scales too small to be resolved with {\sc{trace}} 
(Wedemeyer-B\"{o}hm~et~al.~2007, Jefferies~et~al.~2006).  

Another interesting area incorporating oscillatory phenomena is the work concerned with Evershed flows (see Solanki~2003 for a 
review). Pioneering work by St.~John~\cite{StJ13} 
states that the Evershed velocity flow in sunspots induces plasma behaviour which is detectable both in the photosphere and 
chromosphere. This flow of plasma is readily observed in H-$\alpha$ wings and has an associated flow speed of up to 20~km/s.

To date, there has been a multitude of observations undertaken to search for 
oscillations in the upper solar atmosphere (chromosphere, transition region and corona). However, very little work has been carried out 
to search for high-frequency oscillations in the photosphere, where the building blocks of wave generation and propagation may exist.

Van Noort \& Rouppe van der Voort~\cite{vanN06} have verified the existence 
of highly dynamic structures, including propagating waves with velocities exceeding 200~km/s, in the chromosphere. This, coupled with 
the detection of fast fluctuations of H-$\alpha$ emission, from a flare kernel on timescales of 0.3-0.7~s, by Wang~\cite{Wan05} 
demonstrates the need for high-cadence solar imaging. 
Furthermore, using typical chromospheric plasma parameters, the cooling time of the chromosphere is 
approximately 1~s (Heinzel~1991). Therefore, in order to study such rapid atmospheric variations, it is necessary to implement 
high-cadence imaging techniques. When searching for rapid solar variability, in particular, high-frequency oscillations, there 
are many obstacles which first must be overcome. The search for 
rapid, often low-amplitude intensity variations requires a suitably high cadence to satisfy the Nyquist parameter, a highly sensitive 
camera system providing accurate, sustained frame rates, good seeing and a minimally weakened signal caused by the convolution of 
wave perturbations with a wide response function. 

Here we report intensity oscillations detected within the lower solar atmosphere using a new camera system developed at Queen's University 
Belfast -- the {\sc{rapid dual imager~(rdi)}}. In \S~2 we provide a brief background to the {\sc{rdi}} system used 
for our observations, which are described in detail in \S~3. In \S~4 we discuss the methodologies used during 
the analysis of the observations and the search for reliable oscillatory signatures. A discussion of our results in the context of 
Evershed oscillations and penumbral waves is in \S~5, and finally, our concluding remarks are given in \S~6.

\section{Rapid Dual Imager}
\label{RDI}
 
The {\sc{rdi}} was developed as a follow on to the highly successful {\sc{SECIS}} (Phillips~et~al.~2000) camera system and consists of 
two Basler A301b cameras -- one `master' and one `slave' channel -- connected to a custom built PC, and controlled 
by software designed and developed by 4C's of Somerset, UK. The cameras have a 502 $\times$ 494 pixel$^2$ CCD, with a 
square pixel size of 9.9 $\mu$m and can operate at a maximum speed of 80 fps. Their lightweight, robust design and C-mount housing 
make them ideal for a portable system such as {\sc{rdi}}. The camera mounts act as passive heatsinks, ensuring that the cameras are 
kept within the recommended operating temperatures.

Each camera is controlled via a PCI interface card. A lossless compression algorithm loaded into an on-board 
field-programmable gate array increases the data throughput rate to the PC. The cameras are attached to the control cards via 
5-metre-long high-bandwidth cables which carry data, command signals (via a standard RS-232 serial link) and power, and
are synchronized via a ribbon connecting the interface cards. The master card's clock is set from the PCI bus clock; the 
slave clock is in turn synchronized to the master clock so that both channels are truly synchronous. {\sc{rdi}}'s high rate of data recording 
is made possible using an IDE RAID array of two pairs of `striped' disks, each pair acting as one data storage area, assigned to a 
particular camera. While images are streamed to two preallocated files, sample images from the stream are displayed on-screen to allow 
the observers to check factors such as image quality and seeing conditions.

The preallocated space is filled at a semi-compressed rate of 195~kB per image per camera. For maximum window size at a frame rate of 
20~Hz, this corresponds to an uninterrupted data run of over 7~hr using the full available 200~GB of hard-disk space. The data are then 
converted to FITS format using the built in conversion software and backed up onto tape drives or external hard-drives. {\sc{rdi}} is 
equipped with a multitude of additional features, such as the ability to window specific, user-defined regions of the field of view and 
also to bin multiple groups of pixels together for increased count-rates. {\sc{rdi}} 
remains stationed at the National Solar Observatory, Sacramento Peak as a common-user instrument.

\begin{table}
\begin{center}
\caption{Details of image acquisition.}
\label{lines}
\begin{tabular}{lcccc}
\hline
\hline
  Observation    &  Central     &  Filter   & Height of \\
                 &  Wavelength  &  Bandpass & Formation \\
	         &  (\AA)       &  (\AA)    & (km)	\\
\hline
G-band			&   4305.50     &    9.20			       &  $< 250^{a,b}$ \\
H-$\alpha$ blue wing	&   6561.51     &    $0.21^{c}$ 		       &  $< 200^{d,e}$ \\
\end{tabular}
\end{center}
\footnotesize
$^{a}$: Uitenbroek~\&~Tritschler~\cite{Uit06} \\
$^{b}$: Rimmele \cite{Rim04} \\
$^{c}$: Beckers~et~al.~\cite{Bec75} \\
$^{d}$: Ding~et~al.~\cite{Din01} \\
$^{e}$: Leenaarts~et~al.~\cite{Lee06} \\
\end{table}

\section{Observations}
\label{Obser}

\begin{figure*}
\begin{center}
\includegraphics[angle=0,width=15cm]{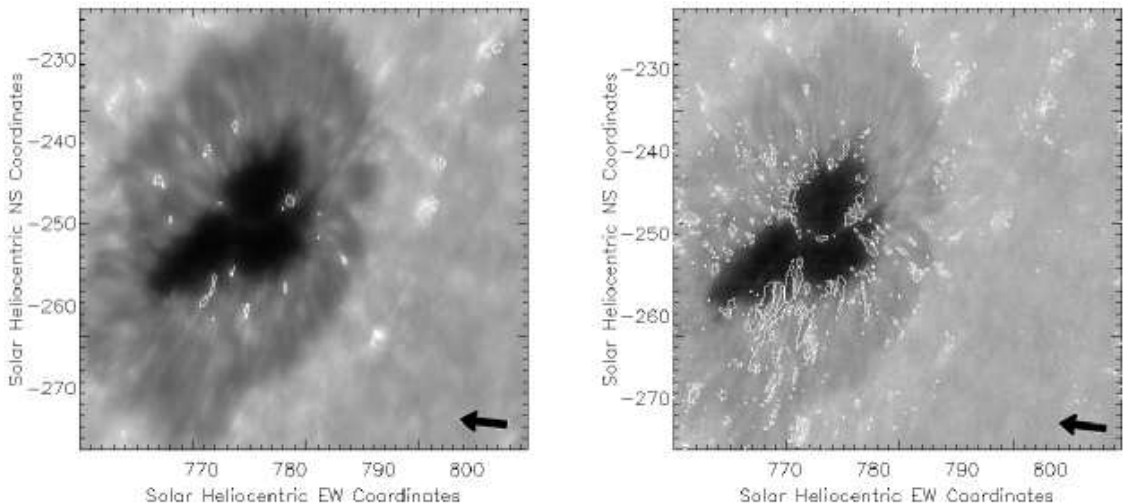}
\caption{\small The G-band (left) and H-$\alpha$ blue-wing (right) field of view. The white contours mark areas where oscillations with a 
26~s periodicity are detected for a particular instant in time. The 
scale is in heliocentric co-ordinates (1~pixel~$=$~$0.1\arcsec$) and all contoured oscillation sites have confidence levels above 99\% 
and abide by the criteria enforced in \S~4.1. The arrow indicates the direction of disc centre.}
\label{rawosc}
\end{center}
\end{figure*}

The data presented here are part of an observing sequence obtained on 2005 August 10, with the Richard B. Dunn Solar Telescope (DST) 
at Sacramento Peak. The optical setup allowed us to image a $50.4\arcsec \times 49.2\arcsec$ region surrounding active region NOAA~10794 
complete with solar rotation tracking. The active region under investigation was located at heliocentric co-ordinates 
($770\arcsec$,$-254\arcsec$), or S12W56 in the solar NS-EW co-ordinate system. A Zeiss universal birefringent filter 
(UBF; Beckers~et~al.~1975) was used for H-$\alpha$ blue-wing (H-$\alpha$ core - 1.3\AA) imaging with one of the {\sc{rdi}} 
CCD detectors. In addition, a G-band filter was employed with the second {\sc{rdi}} camera to enable synchronized 
imaging in the two wavelengths. During the observations presented here, low-order adaptive optics was implemented.

The observations employed in the present analysis consist of 72000 images in each wavelength, taken with a 0.05~s cadence, providing 
one hour of uninterrupted data. The acquisition time for this observing sequence was early in the morning and seeing levels were good with 
minimal variation throughout the time series. Table~1 lists the
wavelengths and filter passbands used in the image acquisition, alongside their approximate heights of formation.
The height of formation of the H-$\alpha$ line profile has long been a topic of key interest among ground-based solar observers. 
Leenaarts~et~al.~(2005) have carried out magneto-convection simulations and show that the formation height of the H-$\alpha$ blue wing 
(core - 0.8\AA) is well under 500~km. Thus, our use of the far H-$\alpha$ blue wing will provide a formation height lower than this, 
forming at a height of 200~km or less (Leenaarts~et~al.~2006). The H-$\alpha$ line, however, is 
very sensitive to solar activity. Indeed, during flares 
the formation height can be pushed into the chromosphere up to a height of approximately 1500~km (Ding~et~al.~2001). During the 
observations discussed here, solar activity was minimal with no flare activity registered within $\pm$~4 days. Utilizing the 
models of Leenaarts~et~al.~(2005, 2006) we therefore conclude that the  formation height of the H-$\alpha$ blue wing in our 
observations is less than 200~km.

A comparison of Figs.~\ref{rawosc}~and~\ref{oscillations} reveals that  many solar features, (sunspot, bright points, granulation 
{\it{etc.}}), are co-spatial and of similar proportions, thus indicating that the H-$\alpha$ blue-wing (core - 1.3\AA) forms at an 
atmospheric height similar to the G-band. Indeed, Uitenbroek~\&~Tritschler~\cite{Uit06}, through the use of 
high-resolution synthetic CH- and CN-band filtergrams, compute the formation height of the G-band to be approximately 200~km. 
Several differences, however, between the two filter bandpasses exist. For example, magnetic elements in the H-$\alpha$ blue wing 
appear to be much brighter and granulation contrast much weaker when compared to simultaneous G-band images. 
Thus, the similar height formation of the two chosen bandpasses, added to their subtle imaging differences, provide us with an 
interesting platform to study processes occurring in the photosphere.
The acquired images have a sampling of $0.1\arcsec$ per pixel to match the 
telescope's diffraction limited resolution in the H-$\alpha$ blue wing to that of the CCD. The optical path to both cameras were identical 
meaning the G-band camera was slightly undersampled. This is desirable to keep the dimensions of the field of view the same for 
both cameras.

\section{Data Analysis}
\label{analy}
\subsection{Initial Image Processing}
\label{raw}

\begin{figure*}
\begin{center}
\includegraphics[angle=0,width=15cm]{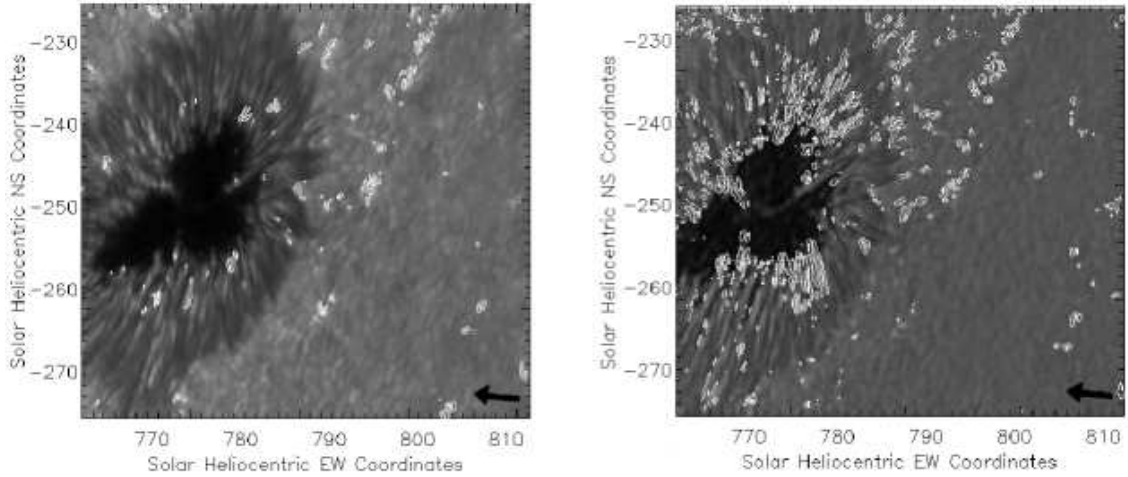}
\caption{\small Speckle reconstructed G-band (left) and H-$\alpha$ blue-wing (right) images. 
The white contours mark areas where oscillations with a 33~s periodicity are detected. The scale is in heliocentric 
co-ordinates (1~pixel~$=$~$0.1\arcsec$) and all contoured oscillation sites have confidence levels above 99\% and abide by the criteria 
enforced in \S~4.1. The arrow indicates the direction of disc centre.}
\label{oscillations}
\end{center}
\end{figure*}

After implementing temporal Fourier analysis on 2400 successive dark images, a moving-pattern noise was revealed, showing up as an intense 
Fourier power with a variable periodicity between 3 and 9~s. This banding was identical in magnitude and periodicity on both cameras. To 
insure this camera induced moving-pattern 
noise did not interfere with the analysis, all periodicities less than 20~s were neglected. Arrays of dark-subtracted and flat-fielded 
data were saved for each camera for subsequent analysis. To 
compensate for camera jitter and large-scale air-pocket motions, 
all data was subjected to a Fourier co-aligning routine commonly available in the {\sc{ssw}} tree of {\sc{idl}}. This routine utilizes 
cross-correlation techniques as well as squared mean absolute deviations to provide sub-pixel co-alignment accuracy. After 5 successive 
co-alignments of the data, the maximum x- and y-displacements, over the entire duration of the dataset, are both less than one tenth 
of a pixel. Since the observing sequence was obtained in the early hours of the morning, 
when image warping is particularly strong, all data were de-stretched relative to simultaneous, high-contrast G-band images. We use a 
$40 \times 40$ grid, equating to a $1.25\arcsec$ separation between spatial samples, to evaluate local offsets between successive 
G-band images. Due to both cameras sharing the same pre-filter optical path, all determined local offsets are applied to simultaneous 
narrowband 
images to compensate for spatial distortions caused by atmospheric turbulence and/or air bubbles crossing the entrance aperture of the 
telescope. The fine destretching grid implemented in this process allows  small-scale seeing conditions, of $1\arcsec$ to $2\arcsec$, 
to be compensated for.

After successful co-alignment and destretching, lightcurves were created for each pixel of each camera before being 
passed into Fast Fourier Transform (FFT) and wavelet analysis routines. While a FFT searches for periodic signatures by decomposing 
the input signal into infinite length sinusoidal wavetrains using a basic exponential function, wavelet analysis utilizes a time 
localised oscillatory function continuous in both frequency and time (Bloomfield~et~al.~2004) and is therefore highly suited 
in the search for transient oscillations. The wavelet chosen for this study is known as a Morlet wavelet and is the modulation of a 
sinusoid by a Gaussian envelope (Torrence \& Compo~1998). Strict criteria were implemented during wavelet analysis to
insure that oscillatory signatures correspond to real periodicities. The first is a test against spurious detections of power that may be 
due to Poisson noise, where the input lightcurve is assumed to be normally distributed (consistent with photon noise) and following 
a $\chi^{2}$ distribution with two degrees of freedom. A 99\% confidence level is calculated by multiplying the power in the 
background spectrum by the values of $\chi^{2}$ corresponding to the 99th percentile of the distribution 
(Torrence \& Compo~1998, Mathioudakis~et~al.~2003). 

The second criterion applied is a comparison of the input lightcurve with a large number (1500) of randomized time-series 
with an identical distribution of counts. The probability, $p$, of detecting non-periodic power is calculated 
for the peak power at each timestep by comparing the value of power found in the input lightcurve with the number of times that the 
power transform of the randomized series produces a peak of equal or greater power. A percentage confidence is attributed to the peak 
power at every time step in the wavelet transform by $(1 - p) \times 100$, such that a high value of $p$ implies that there is no periodic 
signal in the data, while a low value suggests that the detected periodicity is real (see Banerjee~et~al.~2001). 

Our final wavelet criterion is the exclusion of oscillations which last, in duration, less than $1.41$ cycles. This is consistent with the 
decorrelation time defined by Torrence \& Compo~\cite{Tor98}. One can distinguish between a spike in the data and a harmonious periodic 
component at the equivalent Fourier frequency by comparing the width of a peak in the wavelet power spectrum with the decorrelation time. 
From this, the oscillation lifetime at the period of each power maximum is defined as the interval of time 
from when the power supersedes 95\% significance to when it subsequently dips 
below 95\% significance (McAteer~et~al.~2004). The lifetime was then divided by the period to give a lifetime in terms 
of complete cycles (Ireland~et~al.~1999). Any oscillations which last for less than this minimum duration were discarded as they 
may have simply been a spike in the lightcurve.

Four-dimensional maps containing spatial information as well as wavelet power and oscillatory period were saved as outputs of 
wavelet analysis for detected oscillations which lay above the 95\% significance level set by the criteria above.

\subsection{Speckle Reconstructed Data}
\label{speckle}

\begin{figure*}
\begin{center}
\includegraphics[angle=0,width=13.5cm]{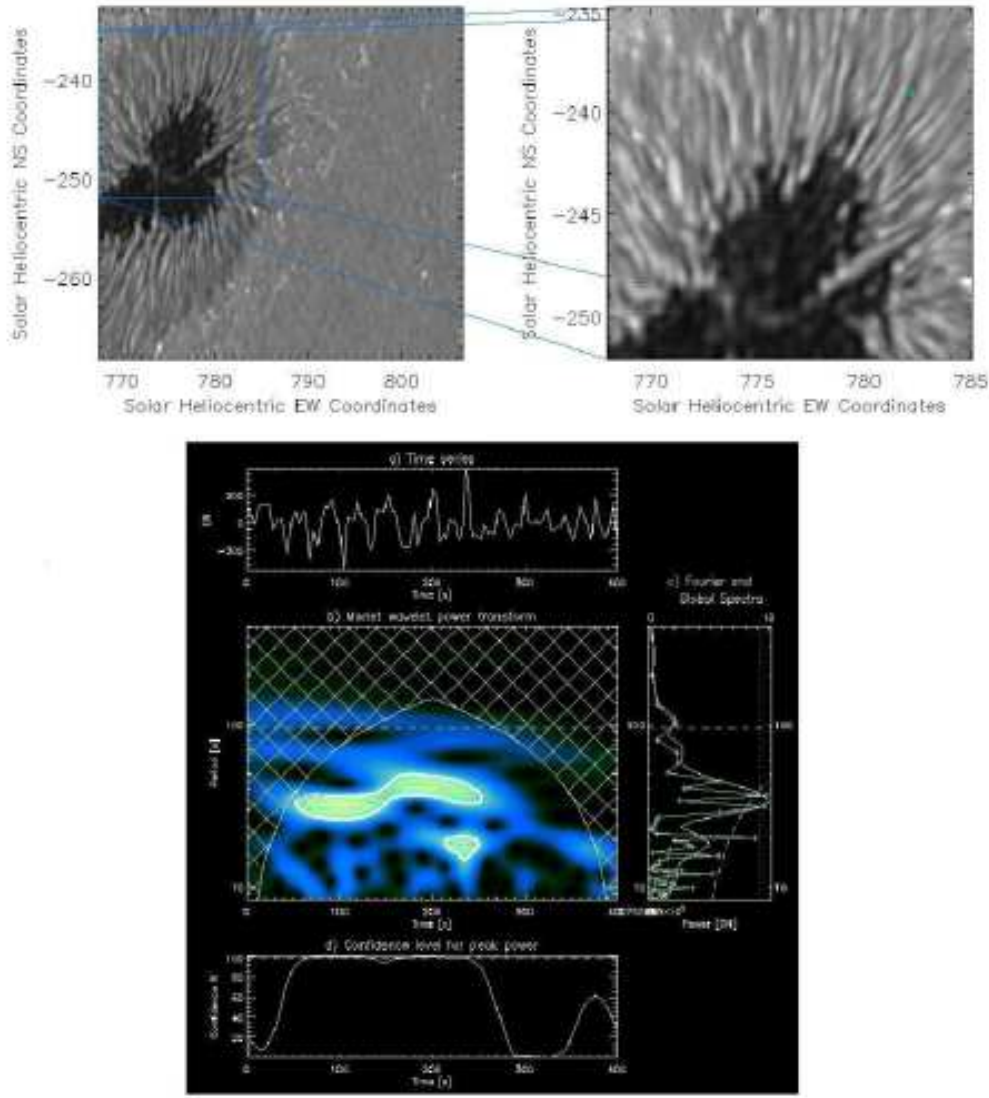}
\caption{\small The top left shows a frame taken from the H-$\alpha$ 
blue-wing datacube followed by a zoomed in portion of the penumbra in the top right. The pixel location from which the lightcurve used 
in wavelet analysis was created is indicated by the cross. The bottom diagram shows the original pixel lightcurve in a). The wavelet 
power transform along with locations where detected power is at, or above, the 99\% confidence level are contained within the contours 
in b). Plot c) shows the summation of the wavelet power transform over time (full line) and the Fast Fourier power spectrum (crosses) 
over time, plotted as a function of period. Both methods have detected a well pronounced 35~s oscillation. The global wavelet 
(dotted line) and Fourier (dashed dotted line) 95\% significance levels are also plotted. The cone of influence (COI), cross-hatched 
area in the plot, defines an area in the wavelet diagram where edge effects become important and as such any frequencies outside the 
COI are disregarded. Periods above the horizontal line (dotted) fall within the COI. The probability levels $(1 - p) \times 100$ as 
discussed in \S~\ref{raw} are plotted in d).  }
\label{wavelet}
\end{center}
\end{figure*}

Small-scale turbulent seeing in the Earth's atmosphere means that even high-order adaptive optics cannot compensate for all 
rapid air motions, and speckle reconstruction is often used as a powerful post-processing routine designed to restore the data to 
diffraction-limited resolution. 
Here we implement the speckle masking method of Weigelt \& Wirnitzer \cite{Wei83}, adapted for solar imaging by von der 
L\"{u}he \cite{vdL93} and further improved by de Boer \cite{deB95}. By observing at a high cadence, the short 
exposure times acquired essentially freeze out atmospheric distortions and maintain signals at high spatial frequencies, albeit with 
statistically disturbed phases (S\"{u}tterlin~et~al.~2001). It is possible to recover the true amplitudes and phases in Fourier 
space by taking a large number of such short exposure images, called a ``Speckle Burst'', and utilizing an elaborate statistical model. 

Eighty raw data frames were used for each speckle reconstruction producing a new effective cadence of 4~s. This provides a Nyquist 
frequency of 125~mHz and is therefore suitable for the search of oscillations with periods longer than 8~s. Typical Fried parameters 
obtained prior to speckle reconstruction were $r_{0}\approx$~10~cm, indicating good post-speckle image quality. However, to strengthen 
the reliability of our detections and to remain consistent with the analysis performed in \S~\ref{raw}, only oscillations with 
periodicities 
greater than 20~s were incorporated into the analysis. The above mentioned dark subtraction was performed identically prior to Speckle 
reconstruction, after which co-aligning and pixel-by-pixel lightcurve analysis was again implemented. Due to the increased spatial 
resolution provided by the implementation of Speckle reconstruction, only 3 successive runs of the co-alignment software was required 
to provide sub-pixel shifts. Again, four-dimensional maps containing spatial information as well as 
wavelet power and oscillatory period were saved as outputs of wavelet analysis for detected oscillations which lay above the 
95\% significance level.

\section{Results and Discussion}
\label{results}

\begin{figure}
\begin{center}
\includegraphics[angle=0,width=8.5cm]{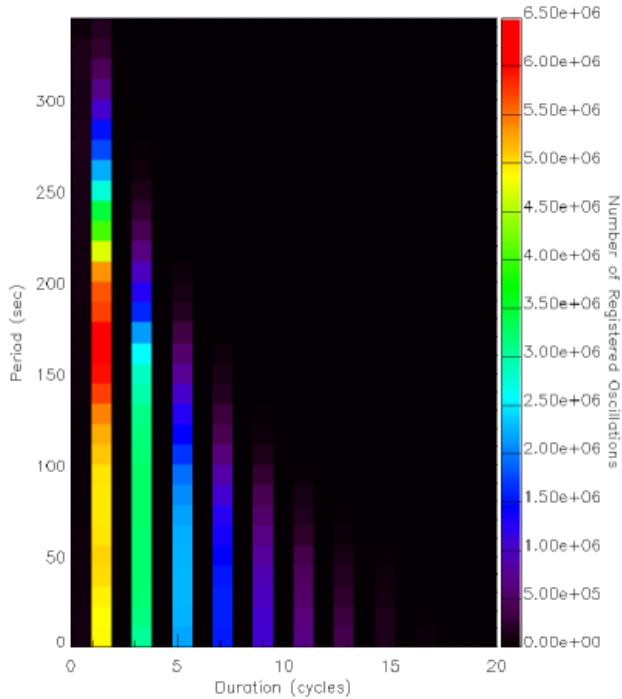}
\caption{\small The duration of oscillations is shown as a function of the period and this plot demonstrates the high occurrence of 
H-$\alpha$ blue-wing oscillatory phenomena. The minimum oscillatory duration 
plotted is $1.41$ cycles as described in \S~\ref{raw}. Each detected period is placed into bins corresponding to its cycle 
duration. For example, a lifetime of 2.65 cycles is placed in the 2--3.99 cycle bin. Cycle duration bins are placed along the x-axis 
beginning with 1.41--1.99, 2--3.99, 4--5.99, 6--7.99 {\it{etc.}}, while the detected period range is placed along the y-axis. Evaluated 
periods are again placed into bins which are 10~s in size beginning with a 0--9.99~s period bin. All oscillations with periodicities 
less than 20~s were not studied in order to fully satisfy the data's Nyquist parameter stated in \S~\ref{speckle}. The colour 
scale represents the number of pixels detecting oscillatory phenomena throughout the entire time series and over the whole field of 
view which are consistent with the criteria enforced in \S~\ref{raw}.}
\label{Halphahist}
\end{center}
\end{figure}

Both processed and speckle reconstructed data reveal signatures of high-frequency oscillations in the G-band and 
H-$\alpha$ blue wing (Fig.~\ref{rawosc}). The well-established 5 minute global oscillation can be seen predominantly in regions 
away from the sunspot (i.e. quiet sun),
and indeed the 3 minute umbral oscillation can be detected. This is consistent with the work of Brynildsen~et~al.~\cite{Bry02} 
who report oscillations with a period of 5 minutes are observed in sunspot umbrae, but with considerably less power than locations 
away from the sunspot. However, the oscillations we will concentrate on are those of much higher 
frequency. We have detected high-Fourier power at periodicities between 20 and 370~s in both the G-band and H-$\alpha$ blue wing, with 
a significant concentration of high-frequency ($> 20$~mHz) activity in the sunspot penumbra (Fig.~\ref{oscillations}). The 
detected oscillations have highly-correlated global wavelet power and Fourier power with the additional support of significance 
levels over 95\%. Figure~\ref{wavelet} shows the detection of a 35~s oscillation originating from a dark penumbral filament within 
images acquired in the H-$\alpha$ blue wing.

\begin{figure}
\begin{center}
\includegraphics[angle=0,width=8.5cm]{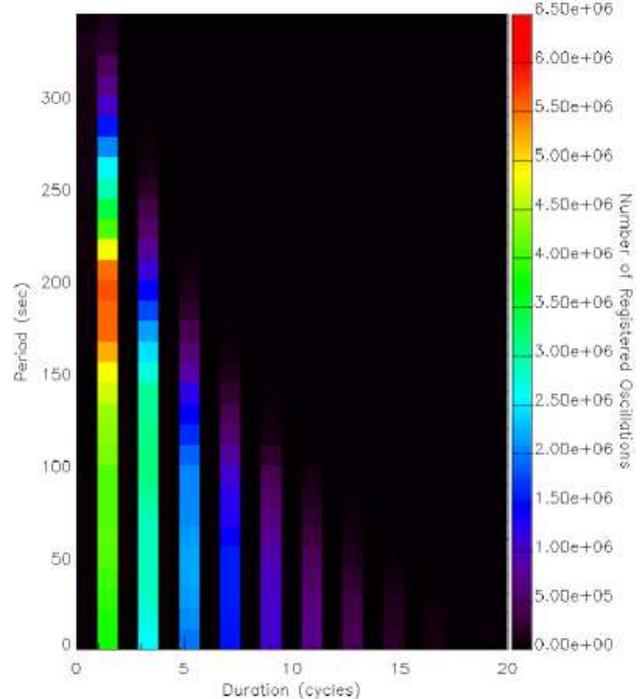}
\caption{\small Same as Fig.~\ref{Halphahist}, but for the G-band. Notice the reduction in detected oscillations 
at higher frequencies when compared to Fig.~\ref{Halphahist}. This can also be viewed via the spatial representation in 
Figs.~\ref{rawosc},~\ref{oscillations}~and~\ref{binned}. }
\label{Gbandhist}
\end{center}
\end{figure}

\begin{figure*}
\begin{center}
\includegraphics[angle=0,width=15cm]{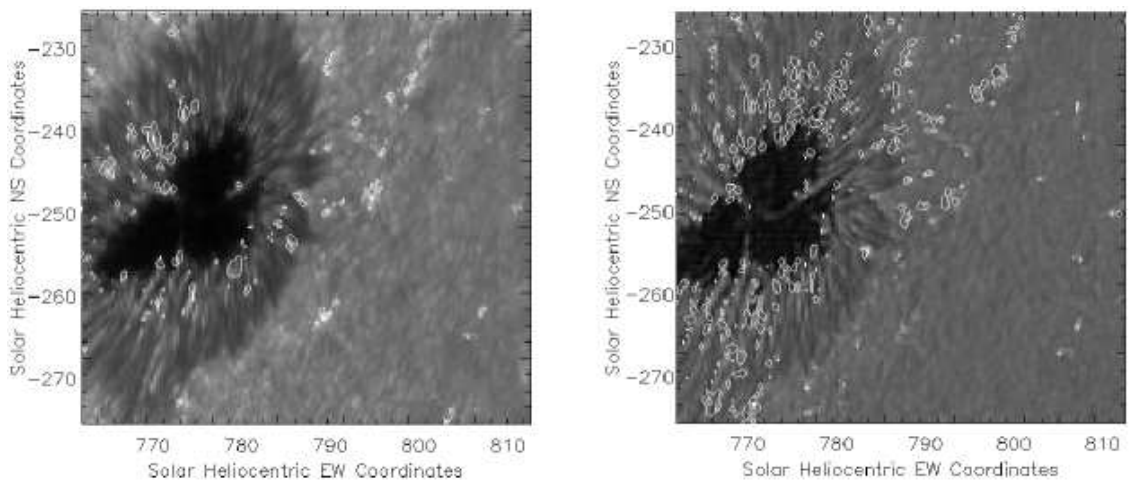}
\caption{\small The speckle reconstructed G-band (left) and H-$\alpha$ blue-wing (right) fields of view with 33~s oscillations 
overplotted. In both images the spatial sampling has been degraded by binning 9 ($3 \times 3$) pixels 
together as discussed in \S~\ref{results}. It is clear to see that penumbral oscillations still dominate in the H-$\alpha$ blue 
wing when compared to simultaneous G-band observations. The scale is in heliocentric co-ordinates and all contoured oscillation sites have 
confidence levels above 99\% and abide by the criteria enforced in \S~4.1}
\label{binned}
\end{center}
\end{figure*}

Figures~\ref{Halphahist}~and~\ref{Gbandhist} show histograms relating to the number of detected oscillations found, in the speckle 
reconstructed data, using the pixel 
by pixel wavelet analysis technique outlined in \S~\ref{raw}. It is clear that a large number of oscillations have been detected, and that 
the occurrence of higher frequency oscillations is greater in the H-$\alpha$~blue~wing. To further test the integrity of our results 
we binned 9 ($3 \times 3$) spatial pixels together and re-computed the behaviour of oscillatory phenomena for both the G-band and 
H-$\alpha$ blue-wing data. This more coarse sampling will help prevent small-scale pixel noise below 3 pixels in amplitude from 
registering as Fourier power during wavelet analysis. From Figure~\ref{binned}, it is clear that the results discussed above are 
unchanged even with coarse sampling.

\begin{figure}
\begin{center}
\includegraphics[angle=0,width=8.5cm]{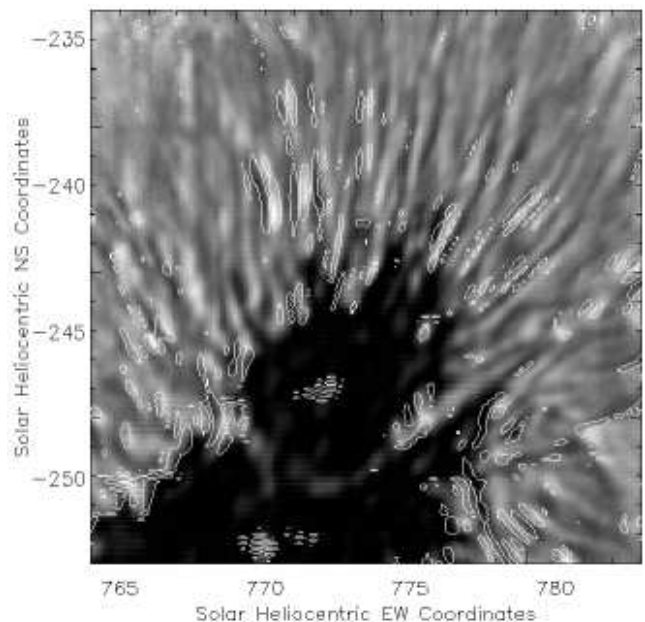}
\caption{\small Speckle reconstructed H-$\alpha$ blue-wing image of the penumbral structure overplotted with locations of detected 
33~s periodicity. The scale is in heliocentric co-ordinates and all contoured oscillation sites have confidence levels above 99\% and 
abide by the criteria enforced in \S~4.1.}
\label{filaments}
\end{center}
\end{figure}

\begin{figure}
\begin{center}
\includegraphics[angle=0,width=8.5cm]{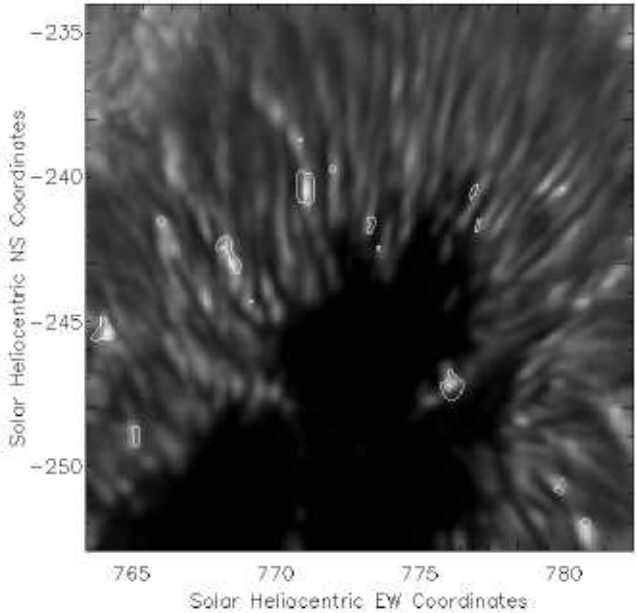}
\caption{\small Speckle reconstructed G-band image of the penumbral structure overplotted with the locations of 33~s oscillations. 
The scale is in heliocentric co-ordinates and all contoured oscillation sites have confidence levels above 99\% and abide by the criteria 
enforced in \S~4.1. Note the reduction in oscillation sites compared with Fig.~\ref{filaments} and how such sites appear to be associated 
with bright penumbral grains.}
\label{brightgrains}
\end{center}
\end{figure}

In this work, we detect significant amounts of oscillatory power originating from bright penumbral grains 
(Figs.~\ref{filaments}~and~\ref{brightgrains}). We propose that this power, detected in both G-band and H-alpha blue-wing observations, 
is closely linked to Evershed flow. A mechanism promoting Evershed flow was proposed by Schlichenmaier~et~al.~\cite{Sch98}, who modelled 
the dynamic 
evolution of a thin flux tube inside the penumbra. A flux tube initially positioned at the magnetopause becomes buoyant due to 
radiative heating and rises. Pressure differences within the loop are created due to radiative cooling at the photosphere which drives 
an outward flow along the flux tube as it rises through the penumbra (Schlichenmaier~et~al.~1998). 
This model also predicts a 
filamentary structure characterized by a hot, nearly vertical upflow of plasma at the footpoint of the filament, which within several 
100~km turns into a horizontal filament. We propose that the oscillations 
detected over bright penumbral grains correspond to the Evershed plasma flow along flux tubes anchored in the photosphere. Indeed, 
Schlichenmaier~et~al.~\cite{Sch98} state that footpoints of flux tubes creating Evershed flow within the 
penumbra could be identified by bright penumbral grains which reiterates our belief that such bright penumbral grain oscillations are 
associated with the Evershed flow.

In order to quantify our results, we determine the percentage of penumbral bright grains, both 
in the G-band and in the H-$\alpha$ blue wing, which demonstrate oscillatory phenomena over a range of periodicities. 
By imposing a minimum bright grain intensity threshold (G-band : penumbral median + 5$\sigma$, H-$\alpha$ blue wing : 
penumbral median + 8$\sigma$) we were able to mask out all regions excluding penumbral bright grains. By overplotting co-temporal 
locations of oscillations in the period range 28--33~s, we were able to determine the percentage of bright grains undergoing oscillatory 
phenomena. After examination of the entire time series, the average number of penumbral bright grains found in the G-band and 
H-$\alpha$ blue wing, respectively, were 17 and 21, using the minimum intensity threshold values stated above. With comparison to the 
location of high-frequency oscillatory power, we determine that 70\% of the G-band bright grains exhibit co-spatial oscillations, while 
79\% of the H-$\alpha$ blue-wing bright grains demonstrate oscillatory perturbations. This reiterates our belief that the oscillations 
detected in penumbral bright grains are indicative of the Evershed flow, as described by Schlichenmaier~et~al.~\cite{Sch98}.

\begin{figure}
\begin{center}
\includegraphics[angle=0,width=8.5cm]{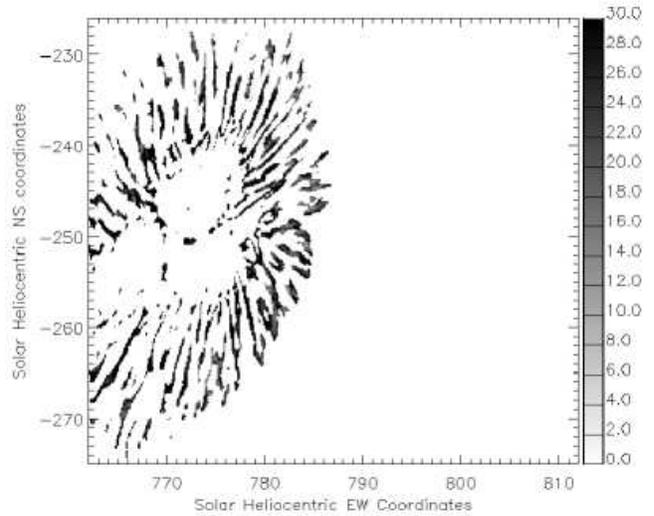}
\caption{\small Average velocity magnitudes of penumbral plasma flows in the H-$\alpha$ blue wing determined from 28~s 
intensity oscillations. The velocities 
displayed represent those of the sunspot only, with all other areas masked out. The axis scales are in heliocentric 
co-ordinates while the colour scale provides an indication to the plasma flow velocity in km/s. Errors 
associated with this diagram may be larger than $\pm~5$~km/s as discussed in \S~\ref{results}.}
\label{velocities}
\end{center}
\end{figure}

From Figs~\ref{rawosc}, \ref{oscillations} and \ref{binned} it is clear  that we detect more oscillatory signatures 
in the H-$\alpha$ blue-wing penumbra compared to the G-band penumbra. Since the formation heights of the H-$\alpha$ blue wing and 
G-band are similar, we are unable to directly interpret these findings based solely on magneto-acoustic oscillations. Instead, 
we believe that this phenomena is a physical signature 
of velocity oscillations in the penumbra, whereby the H-$\alpha$ line profile is periodically shifted due to plasma velocity 
flows. In this regime, wavelet analysis will detect the resulting periodic deviation in intensity as the H-$\alpha$ line profile is 
moved across the filter's 0.21\AA~bandpass. Contrarily, the G-band samples continuum wavelengths and has a broad bandpass (9.2\AA) and 
is therefore only sensitive to intensity rather than velocity variations. After investigating the 
amplitude of intensity variations in the H-$\alpha$ blue-wing penumbra, we find that the average intensity variation detected during 
oscillations, with respect to an average quiescent value for that location on the penumbra, is approximately 11\%. Knowing the tuned 
wavelength of the blue-wing in the H-$\alpha$ profile (6561.51\AA), it is possible to derive the necessary wavelength shift required 
to vary the intensity profile 
by the 11\% observed here. Consulting a solar spectral atlas obtained using the Fourier Transform Spectrometer at the McMath/Pierce Solar 
Telescope, Kitt Peak, Arizona, we have established the corresponding wavelength shift required to produce 11\% amplitude oscillations is 
approximately 0.45\AA. This wavelength shift produces an estimate of the velocity associated with such plasma flows, and we derive this 
velocity to be 20~km/s. Fig.~\ref{velocities} shows the plasma velocities associated with the penumbra as observed in the 
H-$\alpha$ blue wing. It must be noted, however, that since we are performing velocity analysis with narrow-band photometry rather than 
high-resolution spectroscopy, there will be errors in velocity associated with the width of the filter's bandpass. With the 
0.21\AA~bandpass of the UBF, this equates to velocity errors of the order of $\pm~5$~km/s. The use of a solar spectral atlas, for 
determining wavelength shifts from percentage intensity 
perturbations, will introduce some additional errors in the estimated velocities. 
The spectral atlas chosen is produced from disc-centre, quiet sun observations, and as such may differ if near-limb, 
active region data is used. Indeed, an 11\% intensity amplitude could result in very different velocities between disc-centre and 
near-limb observations. Thus, it is imperative to stress potential errors associated with this form of analysis.

It has been known for quite some time that propagating waves exist in the penumbra of active regions. Zirin~\&~Stein~\cite{Zir72} 
found evidence for running penumbral waves (RPWs) in the chromosphere and Giovanelli~\cite{Gio72} who, through the analysis of 
spectroscopic line profiles, discovered associated velocity flows with RPWs of 20~km/s. To date, there has been many detections of 
penumbral velocity flows ranging from 10 to 70~km/s (Brisken~\&~Zirin~1997, Kobanov~et~al.~2006) and 
our detection of 20~km/s velocities is within this range. The photospheric Evershed flow is usually associated with velocities of 
approximately 5~km/s (Bellot~Rubio~et~al.~2003) which is well below our 20~km/s values. However, Schlichenmaier~et~al.~\cite{Sch02} 
provide a model where photospheric Evershed flows may reach velocities up to 
14~km/s, which is comparable with our findings after consideration of the errors involved in our analysis.

Previous observations of Evershed oscillations have detected periods ranging from 8 to 40~min 
(Rimmele~2004, Makarchik \& Kobanov~2001). We note that the penumbral bright-grain oscillations reported here are of much higher 
frequency than those observed to date. The importance of observations in the high-frequency 
domain, has been emphasized by Schlichenmaier \& Solanki~\cite{Sch03} who suggest that hot Evershed upflows may significantly contribute 
to the heating of the penumbra via
the loss of energy through radiative processes as the plasma flows along the flux tubes away from the photosphere. In order to probe 
the viability of such heating mechanisms, it is necessary to combine high spatial and high temporal resolution Dopplergrams of 
non-magnetic lines (to provide a Doppler signal free from magnetic field effects) with high-cadence vector-magnetograms. This will 
allow plasma flow velocities as well as surrounding magnetic field strength to be evaluated, which will enable derivations of the 
penumbral energy flux.

\section{Concluding Remarks}
\label{conc}

We have presented direct evidence of high-frequency waves propagating in an active region, and detected oscillations with 
periodicities ranging from 20 to 370~s with significance levels greater than 95\% due to the rigorous wavelet criteria enforced 
in \S~4.1. More penumbral oscillatory phenomena are found to be located in the H-$\alpha$ blue wing than in the G-band, and we 
conclude that these oscillations are associated with penumbral plasma flows.

H-$\alpha$ blue-wing and G-band oscillations appear to have a strong spatial correspondence with penumbral bright grains. This 
promotes our belief that the 
detected high-frequency oscillations are associated with Evershed flows due to similarities with the model developed by 
Schlichenmaier~et~al.~\cite{Sch98}.

Another aspect to be investigated is the analysis of evolving long-duration oscillatory phenomena. The 
four-dimensional power maps produced during wavelet analysis will be studied for signs of evolution and/or power variability.

\acknowledgements
This work was supported by the U.K. Particle Physics and Astronomy Research Council. DBJ is supported by a Northern Ireland Department for 
Employment and Learning studentship. DBJ additionally thanks NASA Goddard Space Flight Center for a CAST studentship -- in particular Doug 
Rabin and Roger Thomas deserve special thanks for their endless help, support and scientific input. FPK is grateful to AWE Aldermaston 
for the award of a William Penney Fellowship. Observations were obtained at the National Solar Observatory, operated by 
the Association of Universities for Research in Astronomy, Inc. (AURA), under cooperative agreement with the National Science Foundation. 
We are grateful to the anonymous referee for pointing out highly informative references which aided in the interpretation of our 
results. We would also like to thank Kevin Reardon and Gianna Cauzzi for  informal discussions related to the interpretation of the 
findings presented in this paper.
Finally we would like to thank the technical staff at the DST for perseverance in the face of atrocious weather conditions. 
Wavelet software was provided by C. Torrence and G.P. Compo.
\footnote{Wavelet software is available at http://paos.colorado.edu/research/wavelets/.}

~

\bibliography{aa}
\bibliography{submit}

\bibitem[2007]{And07}
Andi\'{c}, A., 2007, Sol. Phys., accepted 
\bibitem[1978]{Ath78}
Athay, R. G., \& White, O. R., 1978, ApJ, 226, 1135
\bibitem[1979a]{Ath79a}
Athay, R. G., \& White, O. R., 1979a, ApJs, 39, 333
\bibitem[1979b]{Ath79b}
Athay, R. G., \& White, O. R., 1979b, ApJ, 229, 1147
\bibitem[2001]{Ban01}
Banerjee, D., O'Shea, E., Doyle, J. G., \& Goossens, M., 2001, A\&A, 371, 1137
\bibitem[1975]{Bec75}
Beckers, J. M., Dickson, L., \& Joyce, R. S., 1975, A Fully Tunable Lyot-\"{O}hman Filter (AFCRL-TR-75-0090; Bedford: AFCRL)
\bibitem[2003]{Bel03}
Bellot Rubio, L. R., Balthasar, H., Collados, M., \& Schlichenmaier R., 2003, A\&A, 403, 47--50
\bibitem[1948]{Bie48}
Biermann, L., 1948, Zeitschrift fur Astrophysics, 25, 161
\bibitem[2004]{Blo04}
Bloomfield, D. S., McAteer, R. T. J., Mathioudakis, M., Williams, D. R., \& Keenan, F. P., 2004, ApJ, 604, 936--943
\bibitem[2007]{Blo07}
Bloomfield, D. S., Solanki, S. K., Lagg, A., Borrero, J. M., \& Cally, P. S., 2007, A\&A, 469, 1155
\bibitem[1997]{Bri97}
Brisken, W. F., \& Zirin, H., ApJ, 478, 814
\bibitem[2002]{Bry02}
Brynildsen, N., Maltby, P., Fredvik, T., \& Kjeldseth-Moe, O., 2002, ESA-SP 506 
\bibitem[1995]{deB95}
de Boer, C. R., 1995, A\&AS, 114, 387
\bibitem[1975]{Deu75}
Deubner, F. -L., 1975, A\&A, 44,371
\bibitem[2001]{Din01}
Ding, M. D., Qiu, J., Wang, H., \& Goode, P. R., 2001, ApJ, 552, 340--347
\bibitem[2005a]{Fos05a}
Fossum, A., \& Carlsson, M., 2005a, Nature, 435, 919--921
\bibitem[2005b]{Fos05b}
Fossum, A., \& Carlsson, M., 2005b, ApJ, 625, 556--562
\bibitem[1972]{Gio72}
Giovanelli, R. G., 1972, Sol. Phys., 27, 71
\bibitem[1991]{Hei91}
Heinzel, P. A., 1991, Sol. Phys., 65
\bibitem[1981]{Hol81}
Hollweg, J. V., 1981, Sol. Phys., 70, 25
\bibitem[1999]{Ire99}
Ireland, J., Walsh, R. W., Harrison, R. A., \& Priest, E. R., 1999, A\&A, 347, 355
\bibitem[2006]{Jef06}
Jefferies, S. M., McIntosh, S. W., Armstrong, J. D., Bogdan, T. J., Cacciani, A., \& Fleck, B., 2006, ApJ, 648, 151--155
\bibitem[2006]{Kob06}
Kobanov, N. I., Kolobov, D. Y., \& Makarchik, D. V., 2006, Sol. Phys., 238, 231-244
\bibitem[2006]{Lee06}
Leenaarts, J., Rutten, R. J., S\"{u}tterlin, P., Carlsson, M., \& Uitenbroek, H., 2006, A\&A, 449, 1209--1218
\bibitem[1960]{Lei60}
Leighton, R. B., 1960, IAUS, 12, 321
\bibitem[1992]{Lit92}
Lites, B. W., 1992, Proceedings of the NATO Advanced Research Workshop on the Theory of Sunspots, 261--302
\bibitem[2003]{McA03}
McAteer, R. T. J., Gallagher, P. T., Williams, D. R., Mathioudakis, M., Bloomfield, D.S., Phillips K. J. H., \& Keenan, F. P., 2003, ApJ, 587, 806--817
\bibitem[2004]{McA04}
McAteer, R. T. J., Gallagher, P. T., Bloomfield, D.S., Williams, D. R., Mathioudakis, M., \& Keenan, F.P., 2004, ApJ, 602, 436--445
\bibitem[2001]{Mak01}
Makarchik, D. V., \& Kobanov, N. I., 2001, IAU Symposium, Vol.~203
\bibitem[2005]{Mar05}
Marcu, A., \& Ballai, I., 2005, PADEU, 15, 103
\bibitem[2003]{Mat03}
Mathioudakis, M., Seiradakis, J. H., Williams, D. R., Avgoloupis, S., Bloomfield, D. S., \& McAteer, R. T. J., 2003, A\&A, 403, 1101--1104
\bibitem[2000]{Phi00}
Phillips, K. J. H., 2000, Sol. Phys., 193, 259--271
\bibitem[2004]{Rim04}
Rimmele, T. R., 2004, ApJ, 604, 906--923
\bibitem[1913]{StJ13}
St. John, C. E., 1913, ApJ, 37, 322
\bibitem[1998]{Sch98}
Schlichenmaier, R., Jahn, K., \& Schmidt, H., U., 1998, A\&A, 337, 897
\bibitem[2002]{Sch02}
Schlichenmaier, R., M\"{u}ller, D. A. N., Steiner, O., \& Stix, M., 2002, A\&A, 381, 77-80
\bibitem[2003]{Sch03}
Schlichenmaier, R., \& Solanki, S. K., 2003, A\&A, 411, 257
\bibitem[1948]{Sch48}
Schwarzschild, M., 1948, ApJ, 107, 1
\bibitem[2003]{Sol03}
Solanki, S. K., 2003, A\&A Review, 11, 153--286
\bibitem[1974]{Ste74}
Stein, R. F., \& Leibacher, J., 1974, ARA\&A, 12, 407
\bibitem[2001]{Sut01}
S\"{u}tterlin, P., Hammerschlag, R. H., Bettonvil, F. C. M., et~al., 2001, ASP Conference Series, Vol. 236
\bibitem[1998]{Tor98}
Torrence, C., \& Compo, G. P., 1998, Bull. Amer. Meteor. Soc., 79, 61
\bibitem[2006]{Uit06}
Uitenbroek, H., Tritschler, A., 2006, ApJ, 639, 525--533
\bibitem[2006]{vanN06}
van Noort, M. J., \& Rouppe van der Voort, L. H. M., 2006, ApJL, 648, 67 
\bibitem[1993]{vdL93}
von der L\"{u}he, O., 1993, Sol. Phys., 268, 374
\bibitem[2005]{Wan05}
Wang, H., Qiu, J., Denker, C., Spirock, T., Chen, H., \& Goode, P. R., 2005, ApJ, 542, 1080--1087
\bibitem[2007]{Wed07}
Wedemeyer-B\"{o}hm, S., Steiner, O., Bruls, J., \& Rammacher, W., 2007, ASP Conference Series, Coimbra Solar Physics Meeting
\bibitem[1983]{Wei83}
Weigelt, G., \& Wirnitzer, B., 1983, Optics Letters Vol. 8, No. 7, 389
\bibitem[1979a]{Whi79a}
White, O. R., \& Athay, R. G., 1979a, ApJs, 39, 317
\bibitem[1979b]{Whi79b}
White, O. R., \& Athay, R. G., 1979b, ApJs, 39, 347
\bibitem[2001]{Wil01}
Williams, D. R., Phillips, K. J. H., Rudawy, P., Mathioudakis, M., et~al., 2001, MNRAS, 326, 428--436
\bibitem[2002]{Wil02}
Williams, D. R., Mathioudakis, M., Gallagher, P. T., Phillips, K. J. H., et~al., 2002, MNRAS, 336, 747--752
\bibitem[1972]{Zir72}
Zirin, H., \& Stein, A., 1972, ApJ, 178, 85

\end{document}